\documentclass[]{algotel}
\usepackage{url}
\usepackage{graphicx}
\usepackage{graphics}
\usepackage{subfigure}
\graphicspath{{Figures/}}
\usepackage{subfig}
\usepackage{epsfig}
\usepackage{wrapfig}
\usepackage[latin1]{inputenc}
\usepackage[english]{babel}

\newcommand{\fenetre}[2]
{$ W_{#1, #2}$}
\newcommand{\prop}[3]
{$#1(W_{#2, #3})$}

\newcommand{\edk}{\emph{eDonkey}}

\newcommand{\udp}{\textsc{udp}}
\newcommand{\ptp}{P2P}

\renewcommand{\marginpar}[1]{}

\author {Lamia Benamara\addressmark{1}{\ }  and Cl{\'e}mence Magnien\addressmark{1}{\ }  }

\title{Removing bias due to finite measurement of dynamic systems: case study on P2P systems}

\address{\addressmark{1}LIP6, CNRS \& Universit{\'e} Pierre et Marie Curie - 4 place Jussieu, 75005 Paris, France -   Prenom.nom@lip6.fr}
          
\begin{document}

\maketitle

\begin{abstract}
Mesurer avec pr{\'e}cision la dynamique des graphes de terrain est une t{\^a}che difficile, car les propri{\'e}t{\'e}s observ{\'e}es peuvent {\^e}tre biais{\'e}es pour diff{\'e}rentes raisons, en particulier le fait que la p{\'e}riode de mesure soit finie. 
Dans ce papier, nous introduisons une m{\'e}thodologie g{\'e}n{\'e}rale qui nous permet de savoir si la fen{\^e}tre d'observation est suffisamment longue pour caract{\'e}riser une propri{\'e}t{\'e} donn{\'e}e dans n'importe quel syst{\`e}me dynamique.

Nous appliquons cette m{\'e}thodologie {\`a} l'{\'e}tude des  dur{\'e}es de sessions et des dur{\'e}es de vie des fichiers sur deux jeux de donn{\'e}es P2P.
Nous montrons que le comportement des propri{\'e}t{\'e}s est diff{\'e}rent : pour les dur{\'e}es de sessions, notre m{\'e}thodologie nous permet de caract{\'e}riser avec pr{\'e}cision la forme de leur distribution.
Par contre, pour les dur{\'e}es de vie des fichiers, nous montrons que cette propri{\'e}t{\'e} ne peut pas {\^e}tre caract{\'e}ris{\'e}e, soit parce qu'elle n'est pas stationnaire, 
soit parce que la dur{\'e}e de notre mesure est trop courte.

\vspace*{-0.4cm}
\end{abstract}

\section{Introduction}

Many systems are naturally dynamic.
For instance in the internet, routers, {\sc as} and/or links between them are created or deleted~\cite{Magnien2009fast};
in peer-to-peer (\ptp{}) networks users join or leave the system~\cite{stutzbach06churn,Saroiu2003,lefessant09}
and exchange different files at different times.
In all these cases, understanding the dynamics of the system is a key issue.
However, accurately measuring these dynamics is a difficult task.
In particular, the fact that the observation window is necessarily finite induces a bias for property characterization~\cite{stutzbach06churn,Saroiu2003}.
Though this bias tends to decrease when the observation window length increases,
it is difficult to quantify it in practice, or know whether it is negligible or not.
\paragraph{}
In this paper, we introduce a new methodology that allows to rigorously determine the minimum observation time required to characterize   
a stationary property 
in real-world dynamic systems.
This methodology is different and complementary to other methodologies existing in the literature~\cite{stutzbach06churn,Saroiu2003,GrossglauserT99},
and has two main advantages. First, it allows to determine if the observation window was long enough for
 a rigorous characterization. Second, it can be applied to any property characterizing the dynamics of a system.
To illustrate its relevance, we apply it to the study of session lengths and files' life duration in two different \ptp{} systems.


\vspace*{+0.2cm}
\footnotetext[1]{We thank Pierre Borgnat, Patrice Abry, Matthieu Latapy, Marcelo Dias de Amorim and Nadjet  Belblida for valuable ideas and comments about
this manuscript.
This work was supported in part by  the French ANR MAPE project,
and by a grant from the {\em Agence Nationale de la Recherche},
with reference ANR-10-JCJC-0202.
This research work is conducted and funded by the European Commission through the EULER project
(Grant No 258307)
part of the Future Internet Research and Experimentation (FIRE) 
objective of the Seventh Framework Programme (FP7).}

\vspace*{-0.1cm}
\section{Methodology}
\label{sec-methodo}

Suppose we start observing a dynamic graph at a time $t$, for a duration $l$.
We denote by \fenetre{t}{l} this observation window.
We are faced with two problems if we want to characterize the graph's dynamics from the observation of \fenetre{t}{l}.
First, $l$ must be long enough for \fenetre{t}{l} to be {\em representative}. Second, even if it is representative,
the fact that $l$ is {\em finite} still induces a bias for property characterization. 
Indeed, events occurring before $t$ or after $t+l$ are not observed, which prevents from characterizing accurately some quantities.
An important point to observe is that the longer the measurement period, the smaller the bias induced. 
\vspace*{+0.1cm}
\paragraph{}
Our methodology addresses these two issues at the same time.
Intuitively, it aims at deciding if the measurement period \fenetre{t}{l} is long enough to 
characterize a given property $P$, i.e. if the bias induced by its finiteness on the observed property is negligible.
If the window \fenetre{t}{l} is long enough, then if we use a longer window \fenetre{t}{l+x}, the observed property does not change:
\prop{P}{t}{l} $=$ \prop{P}{t}{l+x}.
In order to know if a given window is long enough,
we use windows of increasing length \fenetre{0}{l_{1}},\fenetre{0}{l_{2}},  ... \fenetre{0}{l_{n}}, with $l_{1} < l_{2}< ... < l_{n}$. 
By studying how the observed property \prop{P}{0}{l_{1}},\prop{P}{0}{l_{2}}, ...\prop{P}{0}{l_{n}} evolves as a function of $l$,
we determine if it is correctly evaluated or not.
\paragraph{}
\vspace*{-0.1cm}
Finally, an important point is that characterizing a property $P$ only makes sense
if it is stationary, i.e. if $P$ does not evolve while the measurement is under progress.
Notice however that if it is not stationary,
our methodology will not be able to provide a characterization:
the observed property $P$ will not become stable when the observation window length $l$ increases.
If it does become stable,
this means both that \fenetre{t}{l} is long enough,
and that $P$ is stationary.
Notice that, depending on the property studied, other types of bias can occur, see for instance~\cite{stutzbach06churn}, 
including biases coming from the identification of users and their sessions.
We will also rigorously take this into account, see Section~\ref{definition}.

\paragraph{}
\vspace*{-0.1cm}
Here, most of the properties we study are complementary cumulative distributions, 
i.e. for each value $k$, $P_{k}$ is the fraction of all observations values which
are larger than or equal to $k$.

To study how an observed distribution $P$ evolves with the length of the observation window, 
we will first plot the observed distributions \prop{P}{t}{l} for different values of $l$.
In order to confirm more formally the visual observations,
we will also study a statistical indicator which quantifies how close two distributions $P$ and $Q$ are to each other:
the {\em Monge-Kantorovich distance}, or M-K distance~\cite{tryphon09metrics} compares two 
normalized cumulative (complementary or not) distributions $P$ and $Q$.
It is equal to the mean of the distance between the two distributions: 
$MK(P, Q) = (\sum_k |P_{k}-Q_{k}|)/k_{\max}$.

We use this indicator to study how the observed distribution \prop{P}{t}{l} evolves: 
we compute the M-K distance between \prop{P}{0}{l} (with different values of $l$) and \prop{P}{0}{l_{max}},
where $l_{max}$ is the length of the longest observation window for this dataset,
and plot this as a function of $l$. Following~\cite{willinger04variability}, we also study the mean and the standard deviation 
of \prop{P}{0}{l} as a function of $l$.

\section{Data}

In order to show the relevance of our methodology, we use two datasets: the {\em queries} dataset which is a capture of the \udp{} traffic of a large \edk{} server~\cite{hotp2p-serveur}.
It consists of the queries made by users (for lists of files
matching certain keywords, or for providers for a given file),
and of the server's answers to these queries. The measurement lasted for 10 weeks which represents $1$ billion messages, with $89$ million peers and $275$ million files involved.
The {\em logins} dataset consists in a trace of 
the login and logout of peers on the \edk{} network~\cite{lefessant09}.
It contains more than $200$ millions of connections by more
than $14$ millions of peers, over a period of $27$ days. The two datasets are therefore complementary.



\section{Users' session lengths}
\label{sess-length1} 
\subsection{Definition of a session}
\label{definition}
We do not formally know when user sessions begin or end in the {\em queries} dataset,
because there is no notion of session in the \udp{} \edk{} protocol.
Instead, users make stand-alone queries and receive answers from the server.
We therefore have to infer sessions from these queries.

It is natural to consider that
two consecutive queries made by a same user
belong to the same session (whether they are for a same file or not) if the time elapsed between them is short,
and belong to two different sessions if it is long.
The question is then to find an appropriate threshold
for distinguishing between these two cases. Based on the study of the inter-query time distribution (not presented here), we have chosen
to use a threshold of $10\,800$ seconds, i.e. 3 hours.

\subsection{Characterization of session lengths}

We now apply our methodology to the study of the session length distributions $S$, 
by studying \prop{S}{0}{l} for different values of $l$.

\begin{figure}[h!]
 \begin{minipage}[b]{.48\linewidth}
  \vspace*{-1cm}
  \centering\epsfig{figure=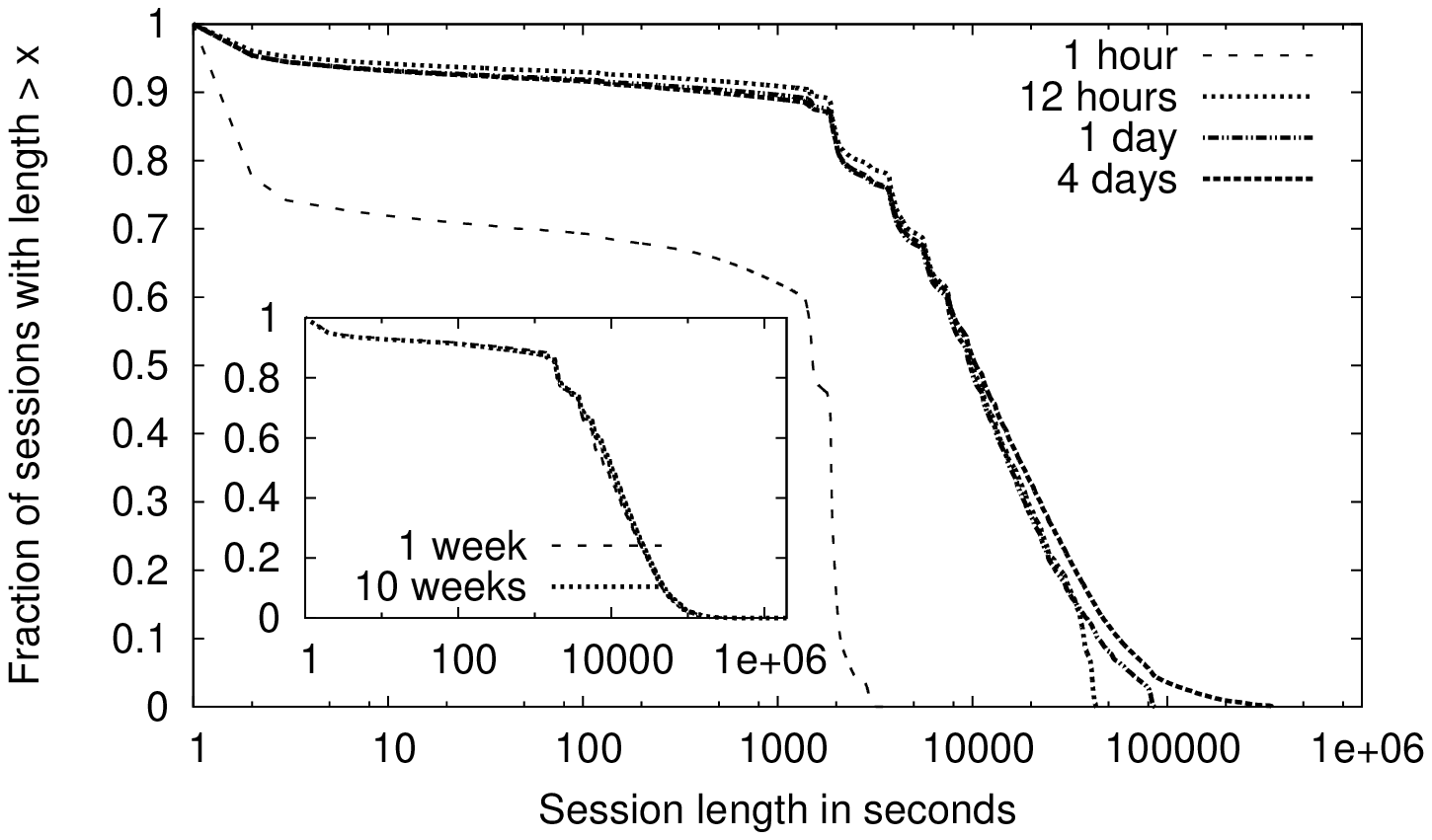,width=.80\linewidth}
  \vspace*{-0.4cm}
  \caption{\small Complementary cumulative distributions of \prop{S}{0}{l} for different observation windows lengths in log-lin scale, for the {\em queries} dataset. \label{fig-1h-7day.t3h}}
 \end{minipage} \hfill
 \begin{minipage}[b]{.48\linewidth}
  \centering\epsfig{figure=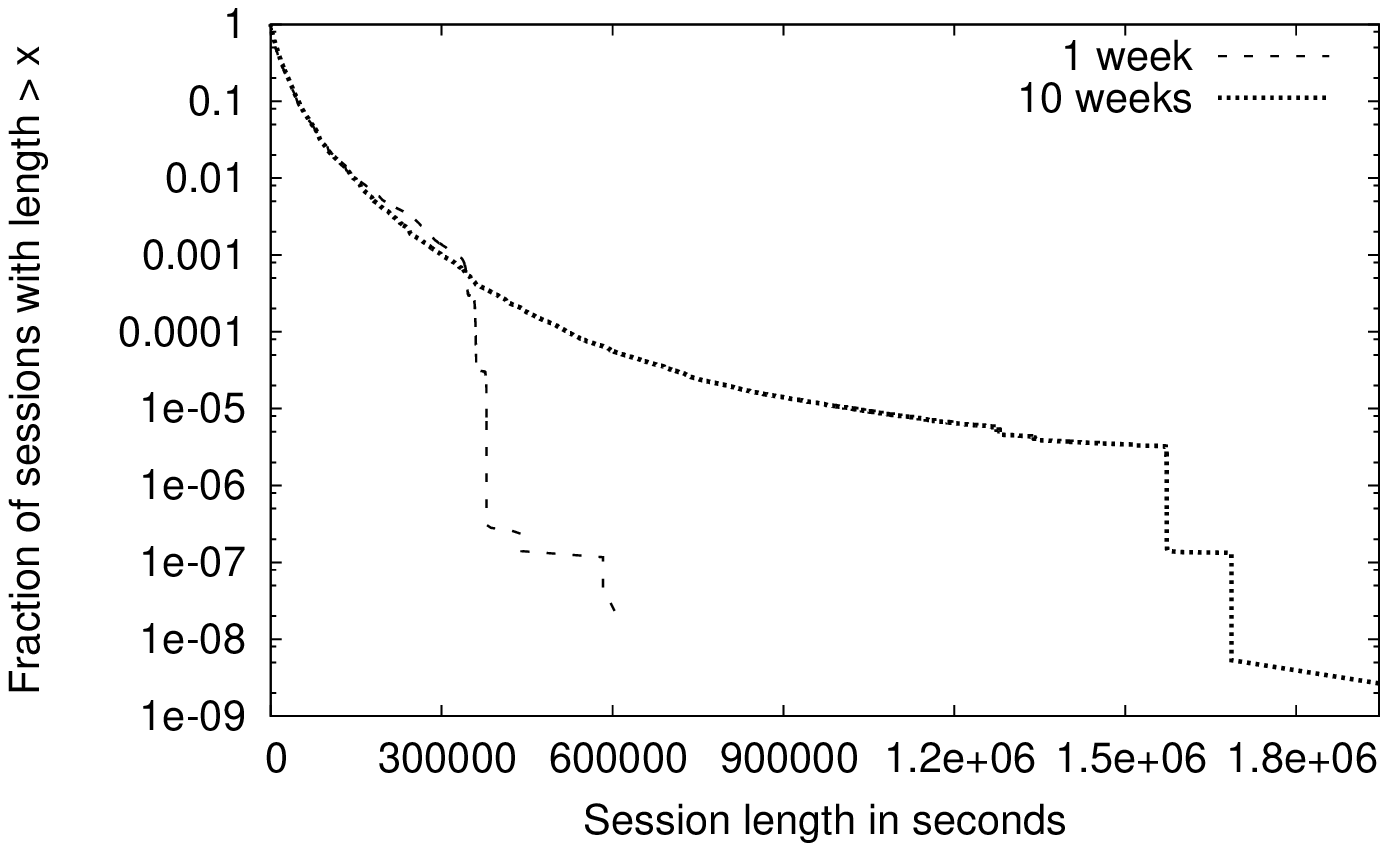,width=.80\linewidth}
  \vspace*{-0.4cm}
\caption{\small Complementary cumulative distributions of \prop{S}{0}{l} for observation windows lengths
    $l=1$ week and $l=10$ weeks in lin-log scale, for the {\em queries} dataset.\label{fig-1w10w-linlog}}
\end{minipage}
\vspace*{-0.2cm}
\end{figure}


Figure~\ref{fig-1h-7day.t3h} shows the complementary cumulative distribution \prop{S}{0}{l}
for different values of $l$, up to $l = 10$~weeks, for the {\em queries} dataset.
The shapes of these distributions are similar,
with a small fraction of sessions with length smaller than $2\,000$ s,
and an approximately linear shape between $2\,000$ s and $100\,000$~s.
However, when $l\leq 1$ day,
the distributions exhibit a clear cut-off.
This is not the case anymore for $l \geq 4$ days:
the tail of the distribution flattens
after a bend occurring close to $100\,000$ s ($\sim$ 28 hours),
and we observe a small fraction of {\em extreme} values after this bend.
For observation windows larger than four days,
the shape of the distribution does not seem to evolve anymore:
the distributions corresponding to $l=1$ week and $l=10$ weeks (presented in the inset)
are very similar to each other and to the one obtained for $l=4$ days.


\begin{wrapfigure}{r}{0.37\textwidth}
  \begin{center}
  \vspace{-0.8cm}
  \includegraphics[width=0.37\textwidth]{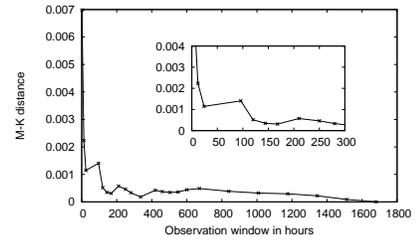}
  \end{center}
  \vspace{-0.5cm}
  \caption{\small MK(\prop{S}{0}{l}, \prop{S}{0}{l_{max}}) as a function of $l$, for the {\em queries} dataset. }
      \vspace{-0.4cm}
  \label{fig-ks-mk}
\end{wrapfigure}

One must be however careful when driving conclusions from a visual examination.
Indeed, if we observe the same plot as the inset of Figure~\ref{fig-1h-7day.t3h}
but with a linear scale on the $x$-axis and a logarithmic scale on the $y$-axis (see Figure~\ref{fig-1w10w-linlog}),
the distributions seem visually strongly different from each other.
However, the distributions are different only for less than $1\%$ of the values, 
which are values after the bend in Figure~\ref{fig-1h-7day.t3h} and
are {\em extreme} values. The fact that the extreme values change when $l$ increase shows
that they cannot be characterized with our methodology, and we leave their study for further work.


To confirm these observations, we study MK(\prop{S}{0}{l}, \prop{S}{0}{l_{max}}) as a function of $l$, presented in Figure~\ref{fig-ks-mk}.
The values observed tend to decrease (with fluctuations) until the observation window 
reaches approximately $150$ hours (6 days and 6 hours).
After this, the value of the M-K distance becomes very small:
this shows that the corresponding distributions are very close to each~other.

We also studied the standard deviation and the mean of \prop{S}{0}{l} as a function of $l$ (not presented here).
We observe that the mean becomes stable once $l$ reaches approximately 1 week, at the same time as the M-K distance.
This confirms that an observation window of one week is long enough to accurately estimate the distribution.
The standard deviation, however, does not seem to converge as the observation window length increases,
confirming that the distribution cannot be {\em fully} characterized.
This is consistent with the distinction between the normal part of the distribution and extreme values.



\begin{figure}[h!]
 \begin{minipage}[b]{.48\linewidth}
  \vspace*{-1.58cm}
  \centering\epsfig{figure=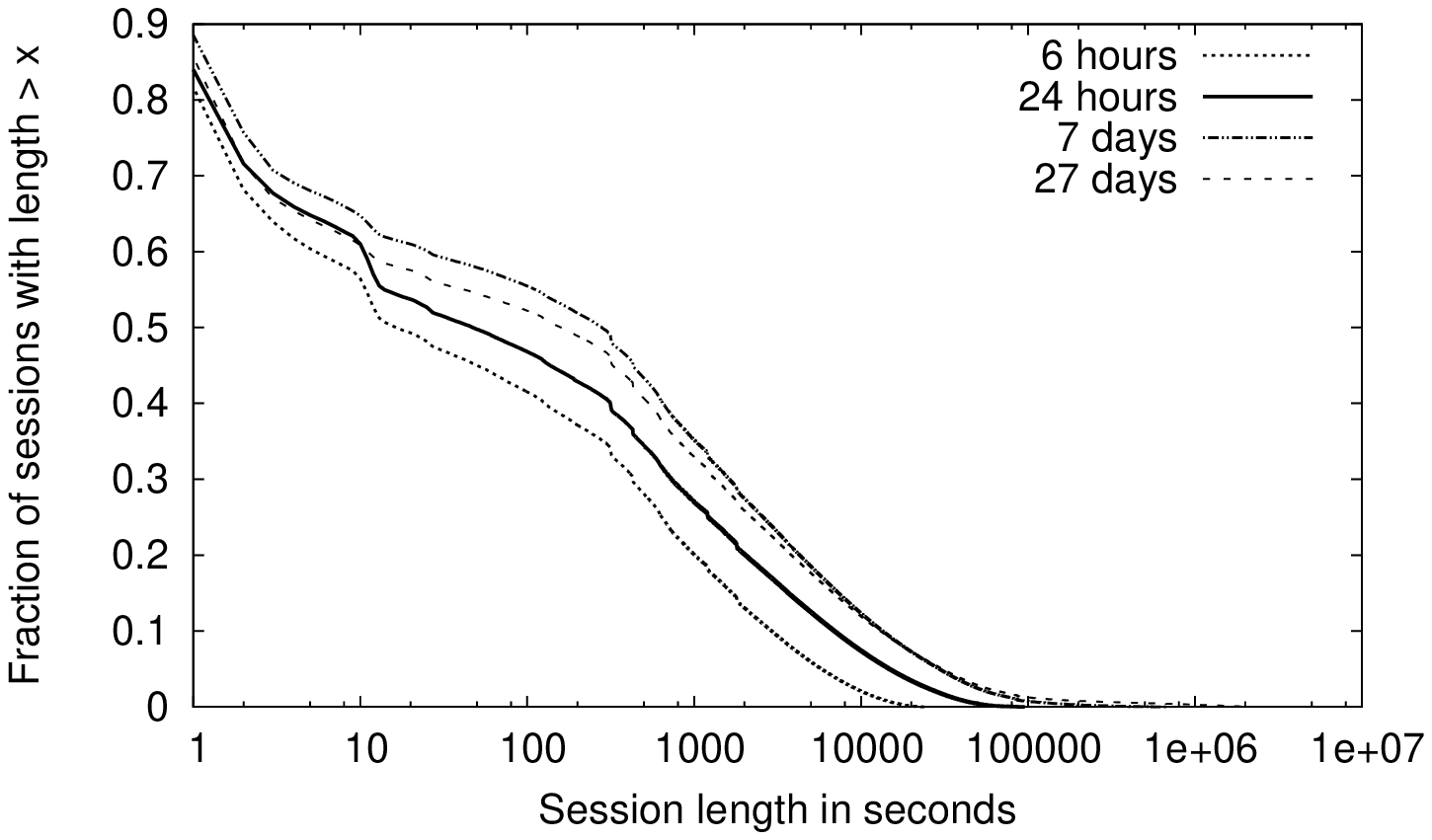,width=.73\linewidth }
  \vspace*{-0.4cm}
  \caption{\small Complementary cumulative distributions of \prop{S}{0}{l} for different observation windows lengths, for the {\em logins} dataset. \label{fig-lefess}}
 \end{minipage} \hfill
 \begin{minipage}[b]{.48\linewidth}
  \centering\epsfig{figure=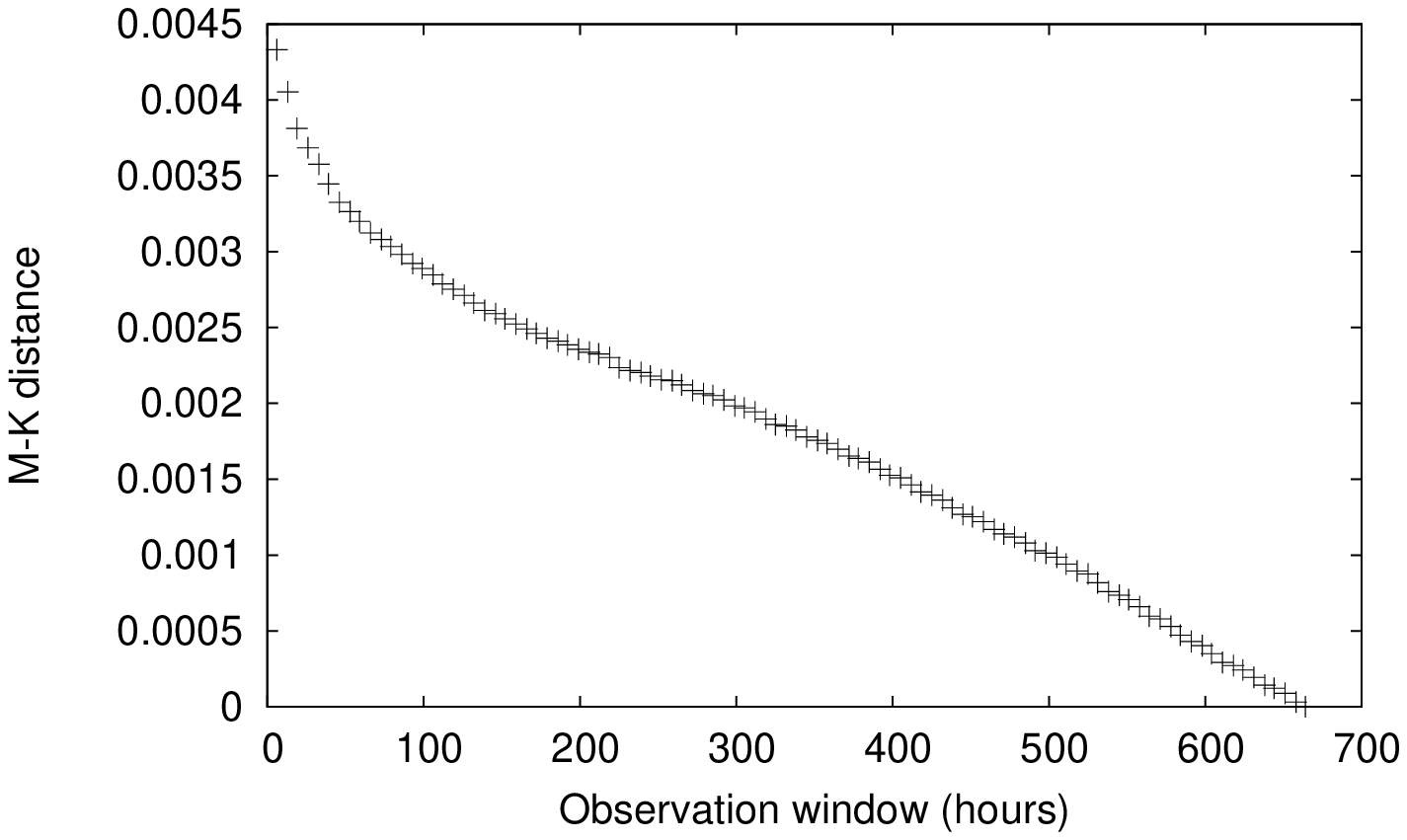,width=.73\linewidth}
  \vspace*{-0.4cm}
 \caption{\small MK(\prop{S}{0}{l}, \prop{S}{0}{l_{max}}) as a function of $l$, for the {\em logins} dataset \label{fig-mk-lefess}}
\end{minipage}
\vspace*{-0.15cm}
\end{figure}

Figure~\ref{fig-lefess} shows the complementary cumulative distribution \prop{S}{0}{l}
for different values of $l$, up to $l = 3$ weeks, for the {\em logins} dataset.
We can see that the shape of these distributions are similar, and get closer to each other as $l$ increases.
However, when we compare these distributions with the M-K distance (see Figure~\ref{fig-mk-lefess}), the values obtained tend to decrease linearly which means that the distributions change at a constant rate.
The values obtained for the mean and the standard deviation also do not stabilize.
Therefore, we can not fully characterize this distribution. We however have confidence that the true shape of the distribution is not far from the one we observed.

\section{Files' lifetime}
We considered two different definitions for a files' lifetime $F$. The first one is the same as for users' sessions lengths: 
we use a threshold and consider that a file is not present in the system if there is no consecutive queries for this file 
separated by less than this threshold. The second definition consists in considering the time interval between the first and the last query for a given file.
In both cases, the shape of distributions \prop{F}{0}{l} (not presented here) evolves strongly with $l$.
We therefore conclude that this property cannot be characterized.
The question which arises is whether this is because this property is intrinsically not stationary or because our measurement period is too short to be able to characterize it.


\section{Conclusion }
\label{sec-conclu}
\vspace*{-0.1cm}
In this paper we introduced an empirical methodology for 
deciding when the bias induced by the finiteness of observation windows in
dynamic systems becomes negligible.
To illustrate the relevance of this approach, 
we applied it to the study of sessions lengths and files' life duration in two different datasets.

We have shown that we can characterize some properties, but not all. 
When a property can't be characterized,  our methodology doesn't allow 
to determine if the observation window shall be increased or not since we don't know the stationarity of the property itself.
It is interesting to note that, for a same dataset, some properties can be accurately characterized, and others not.

\vspace*{-0.1cm}
\bibliographystyle{alpha}
\bibliography{churn}
\label{sec:biblio}


\end{document}